\documentclass[prb,reprint,aps,floatfix]{revtex4-1}
\usepackage{amssymb,graphicx,color,amsmath,bm} 
\usepackage[bookmarks, colorlinks=true, breaklinks]{hyperref}  
\hypersetup{linkcolor=blue,citecolor=blue,filecolor=black,urlcolor=blue} 

\usepackage{soul}

\newcommand{\bra}{\left< }
\newcommand{\ket}{\right>}

\emergencystretch=\maxdimen
\begin{document}

\title{Why is the electrocaloric effect so small in ferroelectrics?}
\author{G. G. Guzm\'{a}n-Verri$^{1,2}$ and P. B. Littlewood$^{3,4}$}
\affiliation{$^1$Materials Science Division, Argonne National Laboratory, Argonne, Illinois, USA 60439 \\
$^2$Centro de Investigaci\'{o}n en Ciencia e Ingenier\'{i}a de Materiales, Universidad de Costa Rica, San Jos\'{e}, Costa Rica 11501 \\
$^3$Argonne National Laboratory, Argonne, Illinois, USA 60439  \\
$^4$James Franck Institute, University of Chicago, 929 E 57 St, Chicago, Illinois, USA 60637}

\begin{abstract}

Ferroelectrics are attractive candidate materials 
for environmentally friendly solid state refrigeration free of greenhouse gases. 
Their thermal response upon variations of external electric fields 
is largest in the vicinity of their phase transitions, which may occur near room temperature. 
The magnitude of the effect, however, is too small for useful cooling applications even
when they are driven close to dielectric breakdown. 
Insight from microscopic theory is therefore needed 
to characterize materials and provide guiding principles to search
for new ones with enhanced electrocaloric performance. 
Here, we derive from well-known microscopic models of ferroelectricity 
meaningful figures of merit for a wide class of ferroelectric materials. 
Such figures of merit provide insight into the 
relation between the strength of the effect and the
characteristic interactions of ferroelectrics such as dipolar forces. 
We find that the long range nature
of these interactions results in a small effect. A strategy is proposed to
make it larger by shortening the correlation lengths of fluctuations of polarization.
In addition, we bring into question other widely used but empirical figures of merit and 
facilitate understanding of the recently observed secondary broad peak in the electrocalorics of relaxor ferroelectrics.

\end{abstract}

\date{\today}
\maketitle

The thermal changes that occur in ferroelectric~(FE) materials upon the application or 
removal of electric fields are known as the electrocaloric effect~(ECE).~\cite{Crossley2015a, Moya2014a, Alpay2014a, Correia2014a, Valant2012a, Scott2011a, Guo2009a} 
The effect was first studied in Rochelle salt in 1930~\cite{Kobeko1930a} and it's 
the electric analogue of the magnetocaloric effect, which is  
commonly used to reach temperatures in the milliKelvin range. 
The ECE is the result of entropy variations with polarization,
e.g., isothermal polarization of a ferroelectric reduces its entropy while
depolarization increases it.
It is parametrized by isothermal changes in entropy $\Delta S$
and adiabatic changes in temperature $\Delta T$ and it is strongest
near the ferroelectric transition. 

Since the phase transitions occur near room temperature in many FEs, 
the potential for using the ECE for cooling applications is huge:
it could provide an alternative to standard refrigeration technologies based 
on the vapor-compression method in which the running substances are greenhouse gases
such as freon and hydrochlorofluorocarbons;~\cite{Narayan2012a}  
replace the widely used but inefficient 
small thermoelectric cooling devices such as Peltier coolers;~\cite{Moya2014a}
and lead to energy harvesting applications.~\cite{Sebald2014a}
Moreover, developing cooling prototypes based on the ECE~\cite{Narayan2009a, Jia2012a, Gu2013a}
may have several advantages
over those based on the more studied magnetocaloric effect as
the magnetic materials of interest require expensive 
rare-earth elements and large magnetic fields, while
many FEs are ceramics or polymers, which are cheap and can be driven with electric
fields that are easy to generate.

Though promising, a major challenge is that
the magnitude of the ECE remains too small for useful applications:
in bulk FEs, $ \Delta T $ is usually less than a few milliKelvin per Volt and 
$\Delta S$ is usually a fraction of a JK$^{-1}$mol$^{-1}$.~\cite{Moya2014a}
FE thin films exhibit ECEs of about an order of magnitude larger than their
bulk counterparts as they can withstand larger breakdown electric fields.~\cite{Mischenko2006a} 
Thin films, however, have small cooling power because of their small heat capacities.
Ferroelectric polymers have also received considerable attention with ECEs comparable to those of thin films, though
they must be driven at larger electric fields than those of thin films.~\cite{Neese2008a}

In the light of these challenges, it has been recently pointed
out that insight from microscopic theory into the ECE
may contribute to characterize known materials and
provide guiding principles to search
for new ones with enhanced electrocaloric performance.~\cite{Moya2014a}
Here, we provide such insight by deriving meaningful figures of merit 
from well-known microscopic models
of ferroelectricity. 
Our figure of merit 
allow us to set trends across different
classes of FE materials (order-disorder and displacive)
and provides insight into the relation between the magnitude 
of the ECE and the characteristic interactions
of FEs (e.g. dipolar and strain). 
We find that the long-range nature of these interactions produces
trade-offs in the ECE: while they can give rise to high 
transition temperatures (i.e., comparable to room temperature),
they concomitantly give rise to long correlation lengths of polarization 
at finite electric fields, which, as we show here, 
result in a small effect.
We make contact with well-known results derived from Landau theory~\cite{Lines1977a}
and those from heuristic arguments.~\cite{Pirc2011b}
Based on these findings, 
we then study the effects of compositional disorder. 
The purpose of this is twofold: to propose
a strategy to increase the magnitude of the ECE
and to 
model the ECE of so-called relaxor ferroelectrics - 
a widely studied class of electrocaloric materials 
with diffuse phase transitions 
that could provide a broad temperature range of operation in 
a cooling device.~\cite{Pirc2011b, Shebanov1992a,Mischenko2006b,Hagberg2008a,Correia2009a,Lu2010a,Valant2010a, Dunne2011a, Rozic2011a,Pirc2011a,Goupil2012a,Perantie2013a,LeGoupil2014a}
We find that the commonly observed secondary broad peak in the ECE of relaxors~\cite{Hagberg2008a, Perantie2013a} is
expected in any ferroelectric that is deep in the supercritical region of their phase diagram. 
Our results also bring into question
the common practice of defining the electrocaloric strength of a material
as the ratio of the entropy or temperature changes over the change in applied electric field.~\cite{Moya2014a,Valant2012a}

To illustrate the ideas presented above,
we adhere to a simple microscopic model for displacive ferroelectrics
with quenched random electric fields.~\cite{Guzman2013a} 
Such compositional disorder 
is typical of relaxor ferroelectrics such as the prototype PbMg$_{1/3}$Nb$_{2/3}$O$_3$~(PMN) and it arises from
the different charge valencies and disordered location of the  Mg$^{2+}$ and Nb$^{5+}$ ions. 
In the absence of disorder, it is a standard minimal model of ferroelectricity.~\cite{Lines1977a}
With disorder, the model gives a good starting point for the description
of the static dielectric properties of relaxors.~\cite{Guzman2013a, Guzman2015a}

In calculating the ECE in the presence of compositional disorder,
it is important to recall 
that Maxwell relations are not applicable. Maxwell relations are usually invoked in ferroelectrics 
to indirectly determine, for instance, 
adiabatic changes in temperature $\Delta T$ from the variations of the polarization with respect to temperature.~\cite{Valant2012a}
Pure ferroelectrics are in thermodynamic equilibrium, thus Maxwell relations hold. 
Disordered ferroelectrics are not in thermodynamic equilibrium, thus 
Maxwell's relations do not apply. 
This is supported by the recent experimental observation 
that direct measurements of $\Delta T$ in
solutions of the relaxor ferroelectric polymer PVDF-TrFE-CFE 
with PVDF-CTFE were significantly larger than those estimated from their polarization curves.~\cite{Lu2010a}  
Another difficulty is that Landau theory fails
for relaxors. Landau theory is applicable 
away from the region of critical fluctuations of polarization that occur
near the transition point. In conventional ferroelectrics, Landau theory
works remarkably well since this region is narrow.~\cite{Lines1977a} In relaxors, on the other hand,
Landau theory is not expected to hold as the region of fluctuations
is broad. We overcome both of these difficulties by calculating the ECE 
directly from the entropy function, as described in the supplementary material (SM).~\cite{SM}

\begin{figure}[htp]
\centering
\includegraphics[scale=0.6]{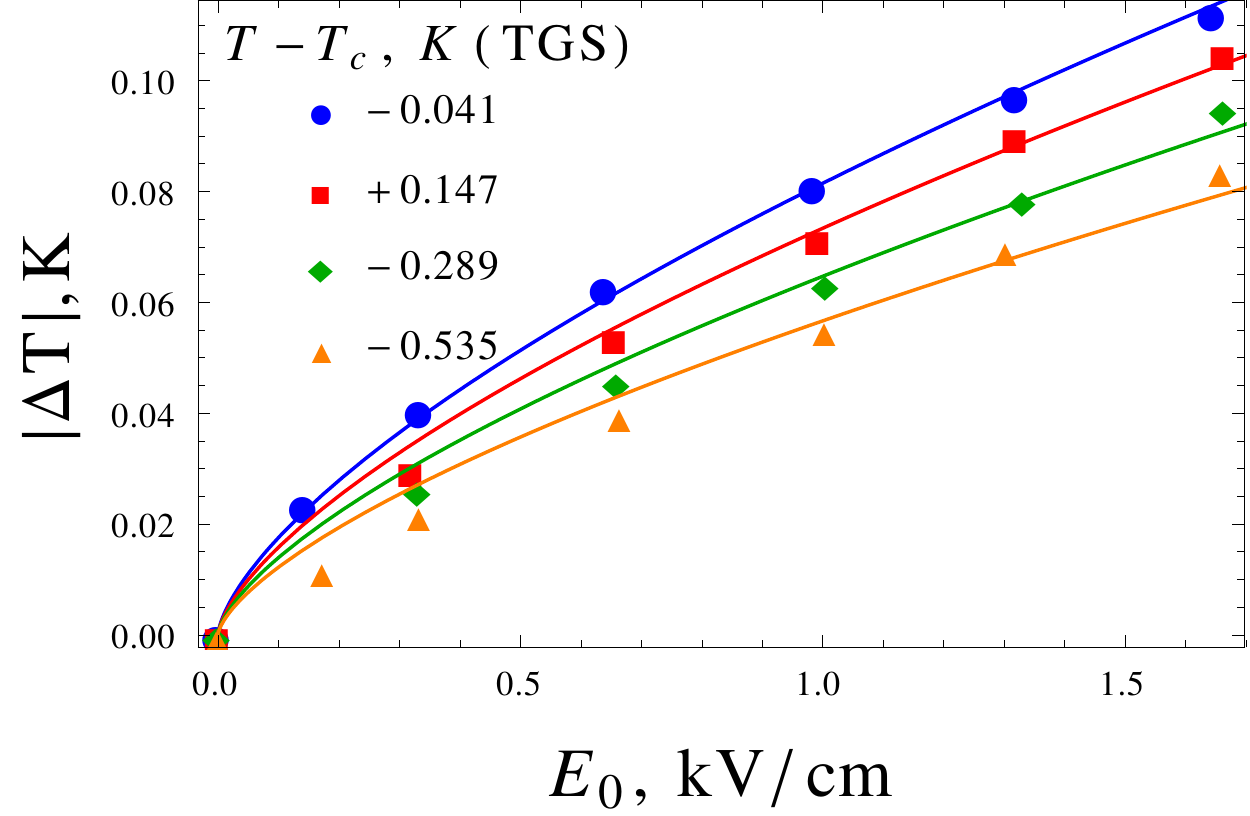}
\caption{Applied electric field dependence of the 
adiabatic changes in temperature near the paraelectric-to-ferroelectric transition~($T_c^0 \simeq 322\,$K) in 
triglycine sulphate~(TGS).
Solid lines correspond to fits to the electric field to the power of $2/3$ as predicted by Eq.~(\ref{eq:fom}).
Data taken from Ref.~[\onlinecite{Strukov1964a}].}
\label{fig:ece_TGS}
\end{figure}

\begin{figure}[htp]
\centering
\includegraphics[scale=0.6]{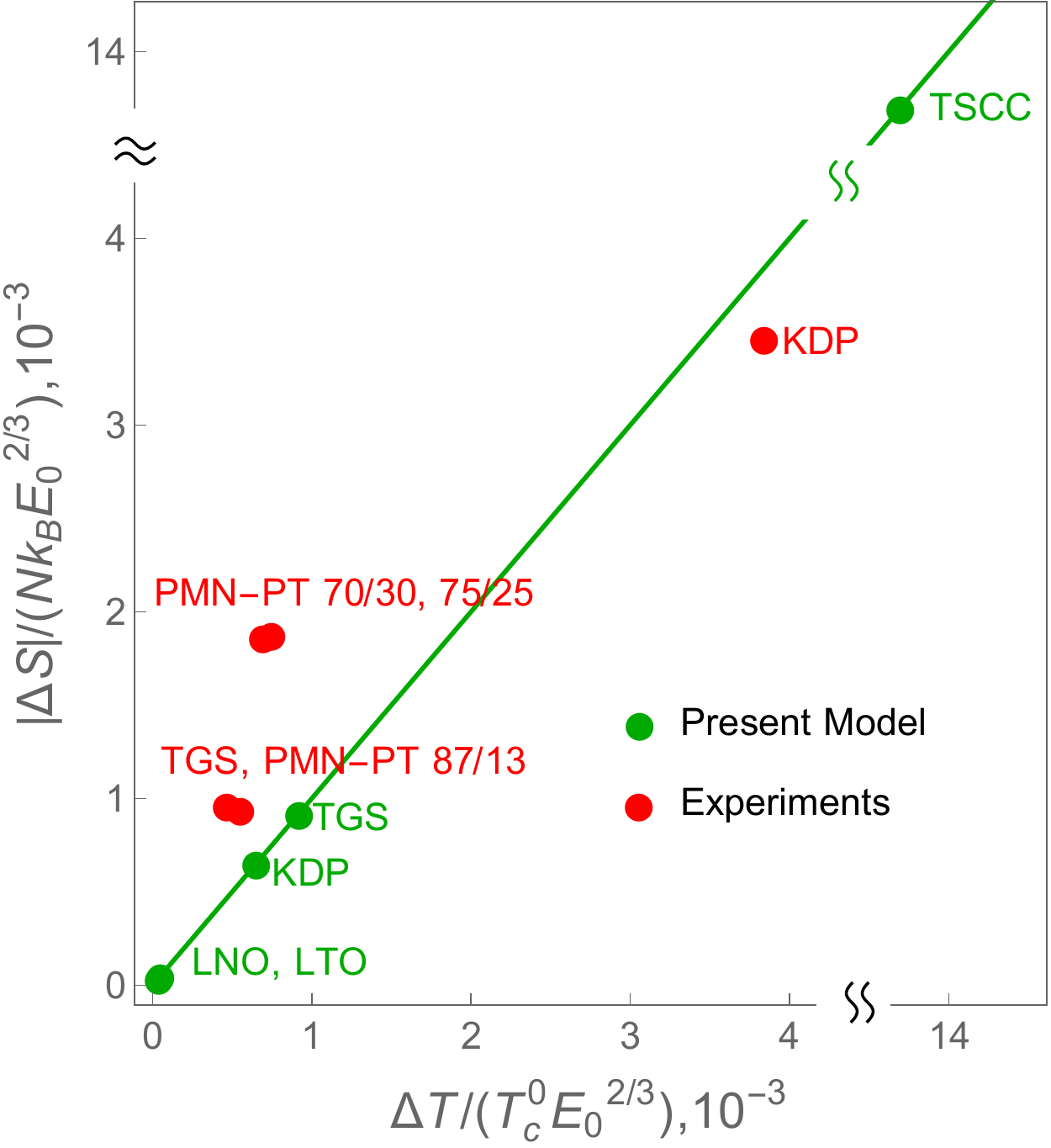}
\caption{   Figures of merit at the paraelectric-to-ferroelectric transition 
for several ferroelectrics. Axes are in units of
(statvolt/cm)$^{2/3}$. 
Here, TSCC = (CH$_3$NHCH$_2$COOH)$_3$CaCl$_2$, KDP = KH$_2$PO$_4$,  TGS = (NH$_2$CH2COOH)$_3$ $\cdot$ H$_2$SO$_4$, 
LNO = LiNbO$_3$, LTO = LiTaO$_3$, and PMN-PT ${1-x}/x$ = (PbMg$_{1/3}$Nb$_{2/3}$O$_3$)$_{1-x}$-(PbTiO$_3$)$_x$.
 Data taken from Refs.~[\onlinecite{Moya2014a}, \onlinecite{Strukov1964a}, \onlinecite{Mackeviciute2013a}].}
\label{fig:ece_fe}
\end{figure}

\begin{figure}
\begin{centering}
\includegraphics[scale=0.32]{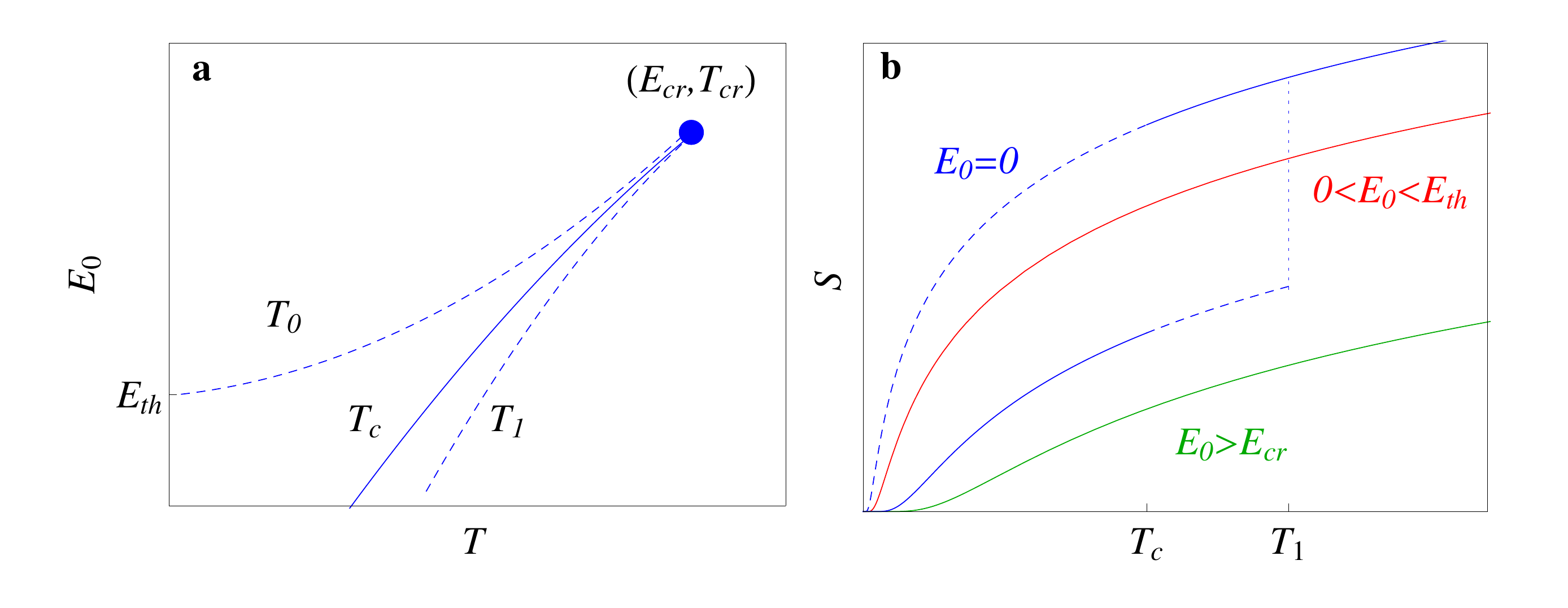}
\end{centering}
\caption{(a) Schematic of the  electric field-temperature phase diagram for ferroelectric with disorder.
The coexistence line~(blue) ends at a critical point $(E_{cr}, T_{cr})$. 
Spinodal curves $T_0$ and $T_1$ indicate the end of the stability of the paraelectric
and ferroelectric states, respectively.
(b) Schematic entropy-temperature phase diagram for various electric field regimes. Solid and dashed lines
correspond to stable and metastable states, respectively. The entropy function for $E_{th} < E_0 < E_{tr}$ is not shown for clarity.}
\label{fig:phase_diagrams}
\end{figure}

\begin{figure*}[htp]
\includegraphics[scale=0.6]{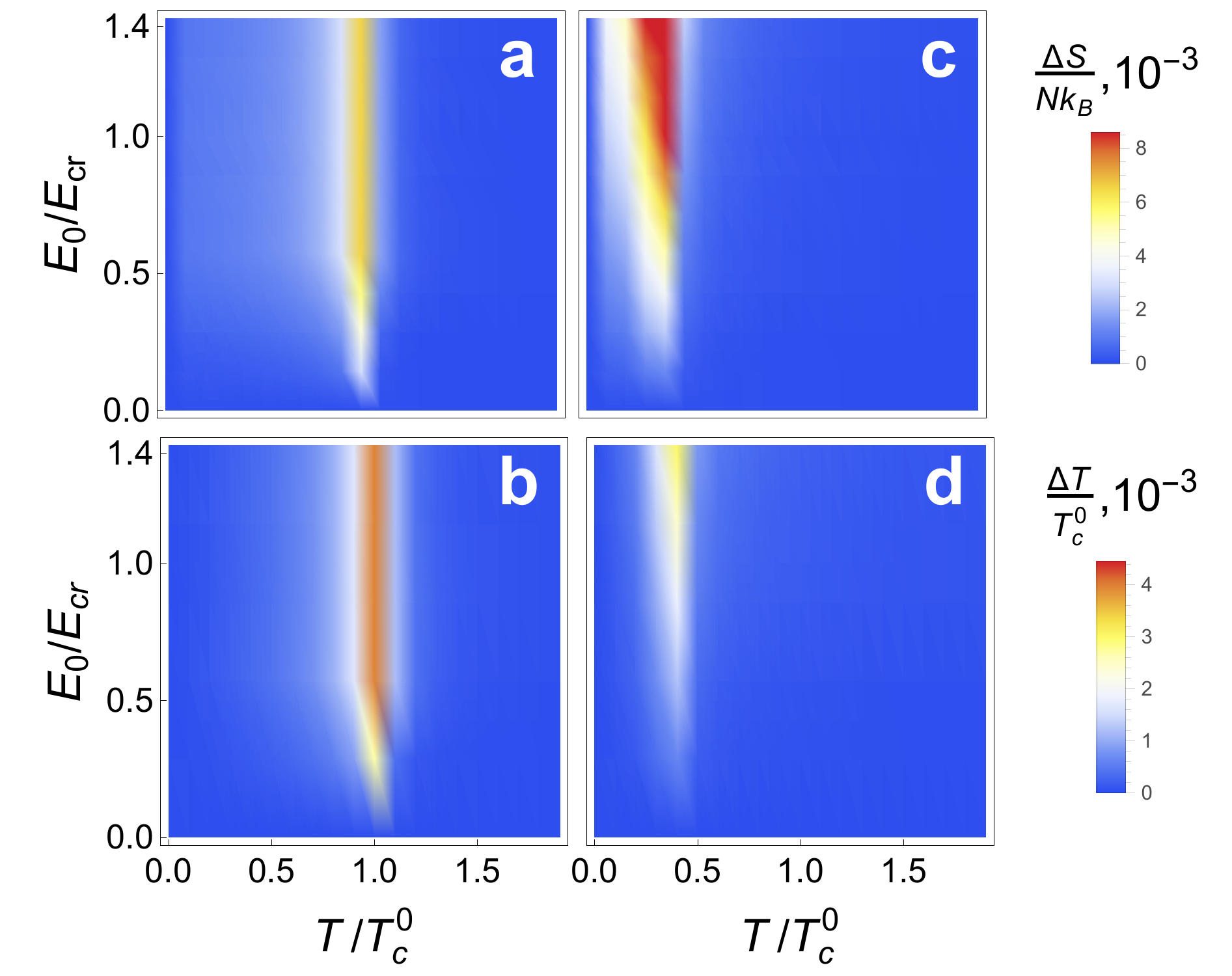}
\caption{Density plots of the adiabatic changes in temperature
and isothermal changes in entropy for (a)-(b) no and (c)-(d) moderate
~($ ( \Delta^2 / v_0 ) / (k_B T_c^0) = 2.0 \times 10^{-2} $) disorder. 
As expected, $\Delta S$ and $\Delta T$ in all cases
are small except at the onset of the ferroelectric transition.
For moderate disorder, $\Delta S$ increases as a result the shortening of the correlation length,
while $\Delta T$ decreases as $T_c$ shifts to lower temperatures. $ v_0  $ is the zone-center Fourier component of the dipolar force.}
\label{fig:ece3D}
\end{figure*}

We first consider the case without compositional disorder. 
In the absence of disorder and no applied electric field, the model gives a second-order paraelectric-to-ferroelectric phase transition
at a transition temperature $ T_c^0 $. 
Consider an isothermal change in entropy $\Delta S(T, E_0)$ near $ T_c^0 $ that results 
from a change in electric field $\Delta E_0 = 0 \to E_0$.
Within a mean field approximation, we find that (see SM),~\cite{SM}
\begin{align}
\label{eq:DeltaS}
\frac{|\Delta S(T, E_0)|}{ N k_B } = \frac{3 \zeta a^2}{8 \pi} \left[ \xi^{-2} ( T,  E_0) -  \xi^{-2} ( T,0) \right],
\end{align}
where $ \zeta$ is a dimensionless coefficient that depends on the lattice structure, $a$ is the lattice constant and 
$N$ is the number of lattice sites.~\cite{Guzman2013a} 
$ \xi ( T, E_0 ) $ is the correlation length of the exponentially decaying fluctuations of polarization at the field $E_0$ and temperature $T$.~\cite{Lines1977a}  
It is given by the  (soft) frequency of the transverse optic phonon mode~\cite{Guzman2013a} and
it diverges as $\xi (0, T ) \propto \left(|T-T_c|\right)^{-1/2}$ at the onset of the FE transition.~\cite{Lines1977a}  
A similar relation is derived for 
the adiabatic changes in temperature $\Delta T(T_1, E_0 ) = T_2 - T_1$  (see SM),~\cite{SM}
\begin{align}
\label{eq:DeltaT}
\frac{\Delta T(T_1, E_0 )}{T_1} =\frac{3 \zeta a^2}{8 \pi} \left[ \xi^{-2} ( E_0, T_2 ) -  \xi^{-2} ( 0, T_1) \right].
\end{align}
Equations~(\ref{eq:DeltaS}) and~(\ref{eq:DeltaT}) relate the ECE to 
the correlation length of a ferroelectric. 
We make contact with known results by noting that
near the FE transition,
$ \left( \xi(T, E_0) / a \right)^{-2} - \left( \xi(T,0) / a \right)^{-2} \simeq (\zeta k_B n)^{-1}  P_s^2(T,E_0) / C_{CW}$ in Eqs.~(\ref{eq:DeltaS}), which
gives the standard results from Landau theory~\cite{Lines1977a} and similar form to those
derived from heuristic arguments.~\cite{Pirc2011b}
$P_s(T, E_0)$ is the polarization at temperature $T$ and at field $E_0$, $C_{CW}$ is the 
Curie-Wiess constant, and $n=N/V$ is the number of lattice points per unit volume $V$.
At the phase transition, $ \Delta S(T_c^0, E_0) $ and $ \Delta T(T_c^0, E_0) $ of Eqs.~(\ref{eq:DeltaS}) and~(\ref{eq:DeltaT}) peak 
as the correlation length at zero field 
diverges~($ \xi ( T_c^0, 0) = \infty $) and 
their magnitude depends on that at finite fields. For ferroelectrics, it is well-known that these tend to be large
due to the long-range nature of the dipole and strain interactions.~\cite{Lines1977a}

We now derive our figure of merit. By evaluating the correlation length at the critical temperature
($ \xi^{-2} ( E_0, T_c^0 )  \propto \left( T_c^0 /C_{CW} \right)^{1/3} \left( E_0/P_s^0 \right)^{2/3}$) 
in Eqs.~(\ref{eq:DeltaS}) and~(\ref{eq:DeltaT}), 
we obtain the magnitude of the ECE in terms of dielectric properties of
a FE  (see SM),~\cite{SM}
\begin{align}
\label{eq:fom}
 \frac{\Delta T( T_c^0 )}{ T_c^0  } \simeq \frac{\Delta S( T_c^0 )}{N k_B } \propto
 \left( \frac{ T_c^0 }{ C_{CW} } \right)^{1/3} \left( \frac{E_0}{P_s^0} \right)^{2/3}, 
\end{align}
where $P_s^0 = P_s(0,0)$ is the saturated polarization at zero field.
The coefficient of proportionality is a dimensionless number~($(1/2)(27/(4\pi))^{2/3} \simeq 0.8$).
Eq.~(\ref{eq:fom}) is our figure of merit.  
 A similar calculation for an order-disorder
FE gives a figure of merit similar to that of Eq.~(\ref{eq:fom}) with two differerences: the ratio $T_c^0/C_{CW}$ is fixed 
to $1/3$ (a well-known result of pure order-disorder models \cite{Strukov1998a}) and a constant of proportionality of 
~($(1/3)^{1/3} (1/2)(27/(4\pi))^{2/3} \simeq 0.6$).

The available data confirm the non-linearity
predicted by our simple model: figure 1 shows that 
$\Delta T \propto E_0^{2/3}$ near the ferroelectric transition of
triglycine sulphate (NH$_2$ CH$_2$COOH)$_3$ $\cdot$ H$_2$ SO$_4$ ). 
Similar scaling laws have been observed in the magnetocaloric effect.~\cite{Franco2010a}
The non-linearity in $E_0$ suggests that 
it is not meaningful to
define $\Delta S / \Delta E_0 $ or $\Delta T / \Delta E_0 $ 
as the electrocaloric strength of an electrocaloric material 
when $ \Delta S $ and $\Delta T$ are measured near or at 
the transition temperature.~\cite{Moya2014a,Valant2012a}  

From Eq.~(\ref{eq:fom}) and data from the literature,~\cite{Landolt-Bornstein}
we calculate our figure of merit for several FE materials. 
The results are shown in Figure~\ref{fig:ece_fe}.
Our model predicts a clear trend: 
order-disorder FEs should display larger ECE than that of the displacive type.
This is a consequence of the shorter correlation lengths that the former type generally
display compared to those of the latter type (order-disorder Curie-Weiss constants are typically about two orders of magnitude smaller than
those of displacive FEs). 
An exception to this rule may exist, however, in the ultraweak displacive FEs 
such as tris-sarcosine calcium chloride~(TSCC). Our predicted figure of merit
is  an order of magnitude larger than any of the FEs considered here as a result of their
exceptionally small Curie-Weiss constants and spontaneous polarization~(which result in shorter correlation lengths).~\cite{Mackeviciute2013a}
The ECE has been recently measured in brominated TSCC compounds, however, we cannot
compare to our model as the measurement was performed near its quantum critical point.~\cite{Rowley2015a}
When Eq.~(\ref{eq:fom}) is contrasted to experiments,~\cite{Moya2014a}  there are clear discrepancies
which we attribute to the mixed order-disorder and displacive character that
most FEs display, and to their large anharmonicities (beyond quartic order).
These differences, though, are not too severe specially when considering the simplicity of the model. 

 We now consider the effects of compositional disorder.
To do so, we consider the minimal model for relaxor ferroelectrics presented in Refs.~[\onlinecite{Guzman2013a}] and [\onlinecite{Guzman2015a}]. Such model includes
dipolar interactions and short-range harmonic and anharmonic forces for the critical modes as in the theory of pure ferroelectrics together with quenched disorder coupled linearly to the critical modes.
In formulating this model, it is important to recall Onsager's result~\cite{Onsager1936a} 
that unlike in the Clausius-Mossotti or Lorentz approximation, dipole interactions alone do not lead to ferroelectic 
order except at $T=0$. Also that pure ferroelectric transitions were understood with the realization~\cite{Cochran-Anderson}  
that they are soft transverse optic mode transitions 
due to dipoles induced by structural transitions 
so that the low temperature phase 
does not have a center of symmetry. 
The first point is not in practice important for pure ferroelectrics which are well 
described by a mean-field theory for the dielectric constant, and is often a first order transition 
only below which the dipoles are produced.~\cite{Lines1977a} But as 
it was shown in Refs.~[\onlinecite{Guzman2013a}] and [\onlinecite{Guzman2015a}],
in relaxors the random location of defects  acts in concert 
with the dipole interactions to extend the region of fluctuations to zero temperatures. Therefore dipole 
interactions must be a necessary part of the model.
The essential physical points in the simplest necessary solution is to formulate a theory which considers thermal 
and quantum fluctuations at least at the level of the Onsager approximation~\cite{Onsager1936a} and random field 
fluctuations at least at the level of a replica theory. The model Hamiltonian and minimal solution are given
in the supplementary material.~\cite{SM}

Following Ref.~[\onlinecite{Guzman2013a}], we
parametrize the quenched random electric
fields by a Gaussian probability distribution with zero mean
and variance $\Delta^2$. \cite{SM} 
Figures~\ref{fig:phase_diagrams}~(a)-(b) show an schematic of the electric field-temperature~($E-T$) 
phase diagram and the entropy function for moderate disorder. 
There are metastable paraelectric states with a stability region that extends to zero temperature.
Ferroelectric states appear as local minima in the free energy at high temperatures
and become stable below a coexistence temperature $T_c < T_c^0$. 
The coexistence line of the polar and non-polar phases ends at a critical point $(T_{cr}, E_{cr})$.
Weak first-order phase transitions are induced for electric fields greater than a threshold field $E_{th}$ as they cross
the region of stability of the metastable paraelectric phase. 
For fields smaller than $E_{th}$, no macroscopic ferroelectric transition occurs 
with a spontaneous polarization. In typical relaxors such as PMN,
$ E_{th} \simeq 2\,$kV/cm and $(T_{cr}, E_{cr} ) \simeq ( 240\,\mbox{K}, 4\,\mbox{kV/cm}  )$.~\cite{Kutnjak2007a}

Figures~\ref{fig:ece3D}(a)-(d) show the electric field and 
temperature dependences of the ECE for zero and moderate 
disorder. In both cases the peaks in  $\Delta S$ and $\Delta T$ 
occur at their corresponding transition temperatures and
increase monotonically with increasing applied field, as expected.
For moderate disorder, however, $\Delta S$ increases provided the applied field is greater than $E_{th}$. 
This increment occurs because of the shortening of the correlation length 
, as indicated by Eq.~(\ref{eq:fom}). Though $\Delta T$
is also affected by the shortening of the correlation length, 
it decreases with disorder as $T_c$ shifts to lower temperatures. 
We emphasize here that we are referering to the correlation length of fluctuations of polarization and 
not to the characteristic nanoscaled size of polar domains of relaxors.~\cite{Shi2011a}

\begin{figure}[htp]
\includegraphics[scale=0.4]{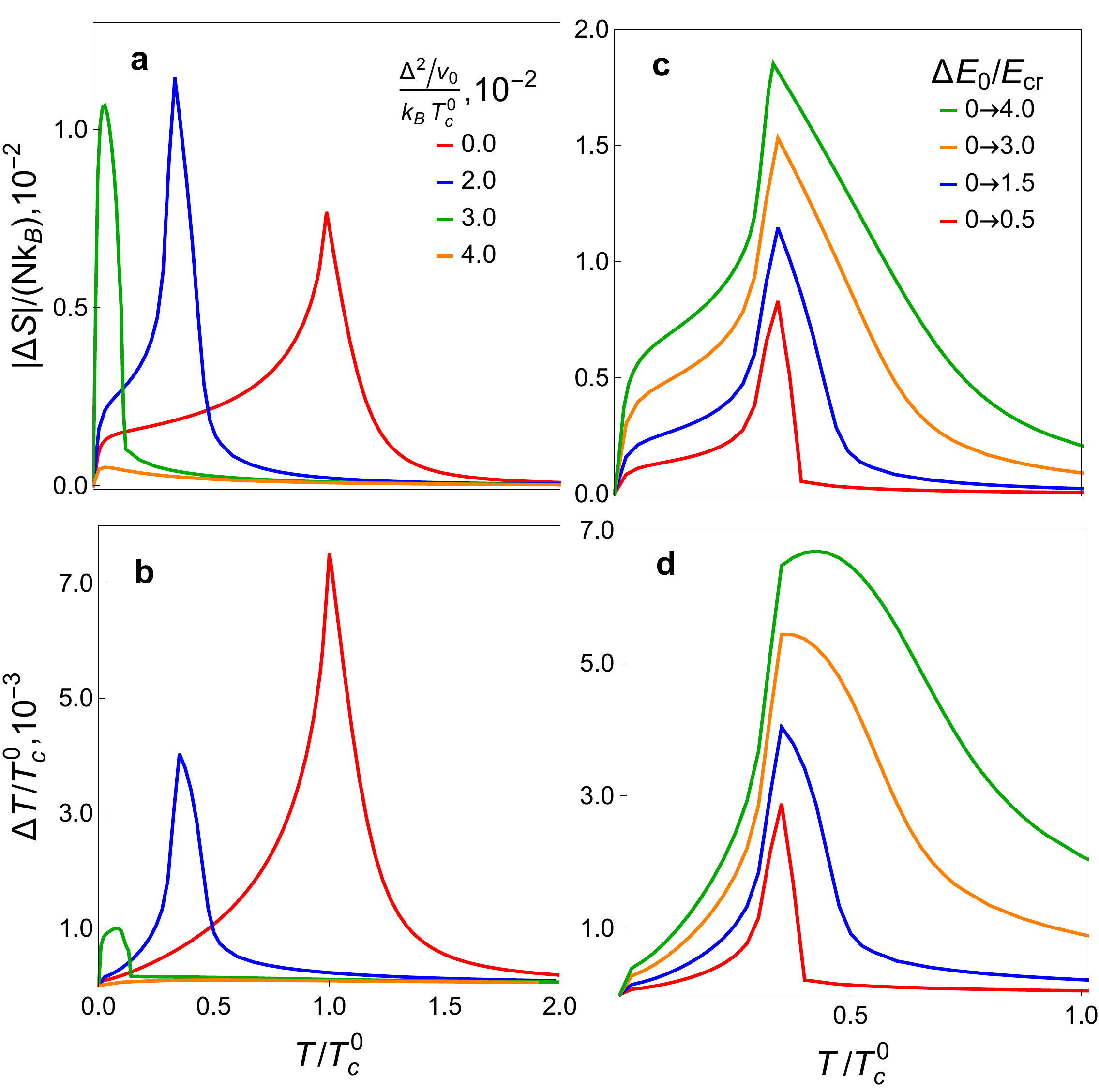}
\caption{(a)-(b) Calculated dependence of the ECE with compositional disorder
for strong changes in the electric field $\Delta E_0/ E_{cr} = 0 \to 1.5$.
(c)-(d) Calculated ECE of a FE with moderate compositional disorder 
~($ ( \Delta^2 / v_0 ) / (k_B T_c^0) = 2.0 \times 10^{-2} $) for several electric field strengths.}
\label{fig:ece}
\end{figure}

Figures~\ref{fig:ece}~(a)-(b) show
$\Delta S$ and $\Delta T$ 
for several disorder strengths
and for a fixed change in the electric field~($\Delta E_0 > 0 \to E_{cr}$).
The increment in $\Delta S$ from weak-to-moderate disorder is clearly shown here
together with the monotonic decrease in $\Delta T$. For strong
disorder, the ECE is weak since the dipoles are pinned by the random
fields, therefore the entropy does not change significantly upon the 
application or removal of electric fields.
Our model is qualitative and fairly good quantitative agreement with the ECE effect observed in typical
relaxors where 
direct temperature measurements in PMN-PT show 
 that a sharp peak in $\Delta T$ shifts to higher temperatures and increases its magnitude with increasing PT concentration 
(up to the morphotropic phase boundary).~\cite{Perantie2013a}

Figures~\ref{fig:ece}~(c)-(d) show
$\Delta S$ and $\Delta T$  
for several electric field strengths and 
for fixed (moderate) disorder. 
As the electric field changes increase, 
a broad peak develops in addition to the usual sharp one at $T_c$. 
Such broad peak in $\Delta T$ is commonly observed in relaxors
and it is usually attributed to nanoscaled polar domains.~\cite{Dunne2011a, Perantie2013a}
However, our model predicts a broad peak occurs in $\Delta T$ already in the absence
of compositional disorder (where there are no polar nanodomains) for very strong fields.
Therefore, the broad peak is simply the expected maximum in the ECE 
for a ferroelectric that is deep in the supercritical regime,  i.e., 
away from the critical point~($ (T, E_0 ) \gg (T_{cr}, E_{cr}) $).
We obtain similar results from Landau theory. \cite{SI} 
Experimentally, this broad peak is not observed 
in conventional ferroelectrics
as their breakdown fields are close to their critical
fields, e.g., $E_{cr} \simeq 10\,$kV/cm~\cite{Novak2013a} and $E_{breakdown} \simeq 14\,$kV/cm~\cite{Landolt-Bornstein} for BaTiO$_3$.

Starting from well-known microscopic models of ferroelectricity, we 
have derived a  meaningful figure of merit for the ECE in a wide class of 
FE materials. When defining a figure of  merit for a caloric effect,
we find crucial to account for the well-known non-linearities 
that occur near the FE transition.  
The large correlation lengths of fluctuations of polarization 
of FEs, result in a small ECE. We predict that ultraweak FEs
such as those of the TSCC-family
should exhibit figures of merit of about an order of magnitude larger
than those of conventional displacive and order-disorder FEs. 
Shortening the correlation lengths results in
larger isothermal changes in entropy, lower transition temperatures,
and smaller adiabatic changes in temperature. 
Broad peaks in the ECE such as those that have been observed in relaxors
are expected for any ferroelectric that can go well into the supercritical region. 
Our work has implications beyond the ECE as a similar analysis could be performed 
for other caloric effects such as
magneto- and mechano-caloric effects, which are currently being studied extensively.~\cite{Moya2014a} 
Our results in combination with other recently proposed figures of merit~\cite{Defay2013a, Moya2015a} should provide guidance for 
characterizing known caloric materials and designing new ones
with enhanced performance.   

\noindent \textbf{Acknowledgements.} We acknowledge useful discussions with Neil Mathur, 
Sohini Kar-Narayan, Xavier Moya, Karl Sandeman, and Chandra Varma.
We thank James Scott for his comments and suggestions on the manuscript. 
Work at Argonne is supported by the U.S. Department
of Energy, Office of Basic Energy Sciences under contract no. DE-AC02-06CH11357.

\newpage
\pagebreak
\widetext
\begin{center}
\textbf{\large Supplemental Material: - Why is the electrocaloric effect so small in ferroelectrics?}
\end{center}
\setcounter{equation}{0}
\setcounter{figure}{0}
\setcounter{table}{0}
\setcounter{page}{1}
\makeatletter
\renewcommand{\theequation}{S\arabic{equation}}
\renewcommand{\thefigure}{S\arabic{figure}}
\renewcommand{\bibnumfmt}[1]{[R#1]}
\renewcommand{\citenumfont}[1]{R#1}

\section{Microscopic Model}

In this section we present the 
microscopic model from which we calculate the
ECE. The model was recently 
proposed in Ref.~[\onlinecite{Guzman2013a}] for relaxor ferroelectrics.
We present it here for the sake of completeness. 
We focus on the relevant transverse optic mode configuration coordinate  $u_i$ of the ions in the unit cell $i$ along the polar axis (chosen to be the $z$-axis).  
$u_i$ experiences a local random field $h_i$ with probability $P(h_1, h_2,...)$ due to the compositional disorder introduced by the different ionic radii and 
different valencies of, say,  Mg$^{2+}$, Nb$^{5+}$, and Ti$^{4+}$ in the typical relaxor (PMN)$_{1-x}$-(PT)$_{x}$. The model Hamiltonian is 
\begin{equation}
\label{eq:Hamiltonian_RFE}
H=\sum_{i}\left[\frac{\Pi_i^2}{2M}+V(u_i)\right]-\frac{1}{2}\sum_{i,j}v_{ij}u_i u_j -\sum_i h_i u_{i} - E_0 \sum_i u_i
\end{equation}
where $\Pi_i$ is the momentum conjugate to $u_i$, $M$ is an effective mass, and  $E_0$ is
a static applied electric field. 
We assume the $h_i$'s are independent random variables with Gaussian probability distribution with zero mean and variance $\Delta^2$.
$V(u_i) =\frac{\kappa}{2} u_i^2 + \frac{\gamma}{4} u_i^4 $ is an anharmonic potential
with $\kappa,~\gamma$ positive constants.
$
v_{ij}/{e^*}^2=
\begin{cases}
3 \frac{(Z_{i}-Z_{j})^2}{|{\bm R}_i-{\bm R}_j|^5} -\frac{1}{|{\bm R}_i-{\bm R}_j|^3}, & {\bm R}_i \neq {\bm R}_j\\
0, & {\bm R}_i = {\bm R}_j,
\end{cases}
$
is the dipole interaction
where $e^*$ is the effective charge and $Z_i$ is the $z$-component of ${\bm R}_i$.
For future use, we denote
$v_{\bm q} = \sum_{i,j}v_{ij} e^{ i {\bm q} \cdot ( {\bm R}_i - {\bm R}_j ) }$
the Fourier transform of the dipole interaction,
$v_{0}~( = 4\pi n {z^*}^2 / 3 )$ 
the ${\bm q} \to 0 $ component of $v_{\bm q}$ in the direction transverse to the polar axis $\hat{ \bm z }$~($ v_{\bm q}$ is non-analytic for ${\bm q} \to 0 $),  
$n$ the number of unit cells per unit volume, and $a$ the lattice constant.

\section{Variational Solution and Entropy Function}

In the present section, we briefly describe the variational 
solution of the problem posed by the Hamiltonian~(\ref{eq:Hamiltonian_RFE})
and obtain an expression for the entropy function. 

We consider a trial probability distribution,
$\rho^{tr} = e^{-\beta H^{tr}} / Z^{tr} $
where $H^{tr} = \sum_i \frac{\Pi_i^2}{2M} + \frac{1}{2}\sum_{i,j} ( u_i - p ) G_{i-j}  ( u_j - p ) - \sum_{i} h_i u_i,$ is 
the Hamiltonian of  coupled displaced harmonic oscillators in a quenched random field,
and $Z^{tr} = \mbox{Tr} e^{- \beta H^{tr} } $ its normalization.
where $\Pi_i$ is the conjugate momentum of the displacement coordinate $u_i$ at site $i$ 
and $M$ is an effective mass. 

We consider random fields~$\{ h_1, h_2, \dots \}$ with Gaussian probability distribution $P(h_1,h_2,...)$ with zero mean and variance $\Delta^2$. 
$p$ is an uniform order parameter: it is the displacement coordinate averaged over thermal disorder first,
and then over compositional disorder,
$p = \int_{-\infty}^\infty dh_1 dh_2 \cdots P(h_1,h_2,...) \, \mbox{Tr} \, \rho^{tr} \, u_i$; 
$\Omega_{\bm q}$ is the frequency of the transverse optic mode at wavevector ${\bm q}$ and it is given by the Fourier transform
of $G_{i-j}$, $ M \Omega_{\bm q}^2 =  \sum_{i,j} G_{i-j} e^{ i {\bm q} \cdot ( {\bm R}_i - {\bm R}_j ) } $. 
We define $  G_{i-j}^{-1} = \left( 1 / N \right) \sum_{\bm q} (M \Omega_{\bm q}^2 )^{-1}  e^{ -i {\bm q} \cdot ( {\bm R}_i - {\bm R}_j ) } $.
$p$ and $\Omega_{\bm q}  $ are variational parameters and are determined by minimization of the free energy
$ F =  \int_{-\infty}^\infty dh_1 dh_2 \cdots P(h_1,h_2,...) \, \mbox{Tr}   \, \rho^{tr} \left[  H  + k_B T  \ln \rho^{tr}  \right] $.

The entropy $S$ is given as follows,
\begin{align}
\label{eq:entropy_1}
\frac{S}{N} &= \int_{-\infty}^\infty dh_1 dh_2 \cdots P(h_1,h_2,...) \, \mbox{Tr} \, \rho^{tr} \, \left( -k_B \ln \rho^{tr}  \right) 
\nonumber  \\
&~~~~~~~~~~~~~~~~~~~~~~~~~
=
\frac{k_B}{N} \sum_{\bm q} \left\{ \frac{ \beta \hbar \Omega_{\bm q} }{ 2 } \coth \left( \frac{ \beta \hbar \Omega_{\bm q} }{ 2 } \right) -  \ln\left[ 2\sinh \left( \frac{ \beta \hbar \Omega_{\bm q} }{ 2 } \right)  \right]  \right\}.
\end{align}
$S$ depends on temperature $T = (k_B \beta)^{-1}$, the applied static field $E_0$; and the stength of compositional disorder $\Delta$.
The adiabatic changes in temperature $\Delta T$ and isothermal changes in entropy $\Delta S$ presented in the main text
are calculated from Eq.~(\ref{eq:entropy_1}). The correlation length of the fluctuations of polarization, $\xi$,
is given by the frequency of the transverse optic phonon at the zone center, $\xi/a = \sqrt{ (3/4\pi) \zeta / (M \Omega_0^2)  }$.~\cite{SGuzman2013a}

\section{Broad Peak in the ECE of Conventional Ferroelectrics}

In this section, we show that Landau theory
predicts a broad peak in the ECE of conventional ferroelectrics. 
We consider the simple Landau theory of Ref.~[\onlinecite{SNovak2013a}]
for the conventional ferroelectric BaTiO$_3$~(BTO) in an applied field $E_0$ along the $(001)$ axis.
Near the the paraelectric-to-ferroelectric transition of BTO,
the Landau free energy is given as follows,~\cite{SNovak2013a}
\begin{align}
\label{eq:Free_BTO}
 F = \frac{a(T-T_0)}{2}P^2 + \frac{b}{4} P^4 + \frac{c}{6} P^6 + \frac{d}{8}P^8 -P E,
\end{align}
where $P$ is the polarization and $T_0 \simeq 400\,$K is the supercooling temperature. 
The coefficients $a = 1.696 \times 10^6\,$ Nm$^2$C$^{-2}$, $b=3.422\times 10^9\,$Nm$^6$C$^{-4}$,
$c=1.797\times 10^{11}\,$Nm$^{10}$C$^{-6}$, $d=3.214 \times 10^{12}\,$Nm$^{14}$C$^{-8}$ are 
determined from the dielectric susceptibility and heat capacity experiments.~\cite{Novak2013a}
The isothermal changes in temperature $\Delta T = T_2 -T_1$, are calculated self-consistently from
the relation $T_2 = T_1 \mbox{exp}\left[ (a / 2 C ) \left( P^2(E_2,T_2) - P^2(E_1,T_1) \right) \right]$,
where the temperature and electric field dependence of $P$ are determined by the standard minimization
procedure of the free energy~(\ref{eq:Free_BTO}). $ C $ is the contribution from the lattice to the 
specific heat. 

In the electric field-temperature phase diagram of BTO
the spinodal line begins at about $(E_0, T_{c}^0) \simeq (0, 405\,\mbox{K})$
and ends at a critical point $(E_{cr}, T_{cr}) \simeq (10\,\mbox{kV/cm},415\,\mbox{K})$.~\cite{SNovak2013a}
The paraelectric-to-ferroelectric transition is discontinous along the spinodal line and
it is continous at $(E_{cr}, T_{cr})$. BTO is supercritical above the critical point and no transition occurs. 

\begin{figure}[h]
\centering
\includegraphics[scale=0.8]{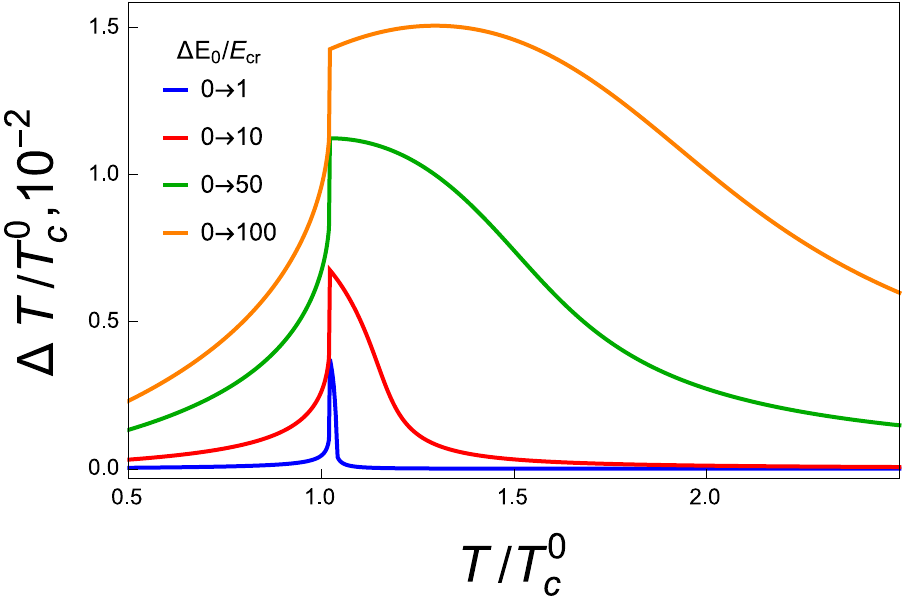}
\caption{Adiabatic changes in temperature, $\Delta T$
in the conventional ferroelectric BTO for several changes in the electric field
strength $\Delta E_0$ applied along the $(001)$ direction. A secondary, broad peak arises in $\Delta T$ in addition
to that at the transition temperature for very large $\Delta E_0$.}
\label{fig:dpBTO}
\end{figure}

Figure~\ref{fig:dpBTO}~(a) shows the adiabatic changes 
in temperature $\Delta T$ for BTO for several changes in the field strength, $\Delta E_0$. 
$\Delta T$ exhibits a single peak at the transition point
for ($\Delta E_0  \lesssim 0 \to 100 E_{cr}$), as expected. 
For larger $\Delta E_0$'s than this, a broad  peak develops. 
Clearly, this peak cannot be observed experimentally
as the required fields are well beyond BTO's breakdown electric field~($\simeq 14\,$kV/cm).~\cite{SLandolt-Bornstein}

\section{Figure of Merit}

The purpose of this section is to derive our figure of merit given in Eq. (\ref{eq:fom}) 
of our manuscript.  To do so we first derive the relation between the 
ECE and the correlation length of fluctuations of polarization (Eqs. (\ref{eq:DeltaS}) and (\ref{eq:DeltaT}) of the main text).

We consider the case of no compositional disorder where a  mean field approximation
is appropriate, as discussed in the main text. 
In the mean field approximation, the free energy per site $F/N$ of 
the Hamtilonian (\ref{eq:Hamiltonian_RFE}) is given as follows, \cite{SBlinc1974a}
\begin{align}
\label{eq:free}
\frac{F}{N} =  \frac{\kappa}{2}\left( p^2 + \sigma  \right)  + \frac{ \gamma }{4} \left( p^4 + 6p^2 \sigma + 3 \sigma^2  \right) 
 - \frac{1}{2}v_0 p^2 - E_0 p  - \frac{ k_B T }{2} -  k_B T \log \left[ 2\pi \sqrt{  M \sigma k_B T  } \right].
\end{align}
Here, $p = \bra u_i \ket$ is a homogenous order parameter and
$ \sigma = \bra  ( u_i  - \bra u_i \ket )^2   \ket $ are mean squared fluctuations. 
Minimization of the free energy with respect to $p$ and $\sigma$ gives the following result,
\begin{align}
\label{eq:mft_eq_RFE_1}
E_0 &= \left( M \Omega_0^2 - 2 \gamma p^2 \right)p,  \\
M  \Omega_0^2 &= \kappa + 3 \gamma \left[ \sigma + p^2 \right] - v_0, \\
\label{eq:mft_eq_RFE_4}
\sigma &= \frac{ k_B T }{ M  \Omega_0^2 + v_0  }.
\end{align}
Here, $  \Omega_0$ is the frequency of the  zone-center soft mode. 
Equations~(\ref{eq:mft_eq_RFE_1})-(\ref{eq:mft_eq_RFE_4})
are self-consistent equations that determine the temperature and electric field dependence of $\Omega_0$ and $p$. 

Equations~(\ref{eq:mft_eq_RFE_1})-(\ref{eq:mft_eq_RFE_4}) 
give a second order phase transition at the critical temperature
$T_c^0 = \left( v_0 - \kappa \right) v_0^\perp / (3 \gamma k_B)$
with a soft mode frequency that follow the Cochran law,
$ M  \Omega_0^2 =  3v_0 (T-T_c)/C_{CW}$
above about $T_c^0$ in the absence of an applied field.
$C_{CW}  =  \left(2 v_0^\perp - \kappa \right)v_0^\perp/( 3 \gamma k_B ) $ is the Curie-Weiss constant.

The soft mode frequency determines the correlation length 
of fluctuations of polarization, $\xi$,~\cite{SGuzman2013a}
\begin{align}
\label{eq:corrlength}
\frac{\xi}{a} = \sqrt{ \frac{3 \zeta v_0  / (4 \pi)}{M \Omega_0^2 }},
\end{align}
where $a$ is the lattice constant and $\zeta$ is a dimensionaless constant. 

The entropy  is given as follows,
$ S(T, E_0) = - \left( \partial F / \partial T  \right)_{p, \sigma, N} = N k_B \left( 1 + \ln \sqrt{ \sigma T }  \right)+$ terms independent of $p,\sigma$ and $T$. Near the FE transition and for weak fields ($ M \Omega_0^2/v_0  \ll 1$),
the entropy is given as follows,
\begin{align}
\label{eq:Sex}
 \frac{S(T, E_0)}{N k_B} = \ln T - \frac{1}{2} \frac{M \Omega_0^2(T,E_0)}{v_0},
\end{align}
where we have ignored terms independent of $T$.
 By substituting Eq. (\ref{eq:corrlength}) into the above equation and considering  changes
in the electric field $\Delta E = E_2 - E_1$, we obtain the following result for the magnitude of the isothermal changes in entropy,
\begin{align}
\label{Seq:DeltaS}
\frac{ \left| \Delta S ( T, E_2, E_1 ) \right| }{N k_B} =   \frac{3 \zeta a^2 }{8 \pi}  \left[ \xi^{-2}(T, E_2) - \xi^{-2}(T, E_1) \right].
\end{align}
This is our desidered result and corresponds to Eq. (1) in the paper. 

We now calculate the adiabatic changes in temperature. 
For adiabatic processes, Eq. (\ref{eq:Sex}) gives
\begin{align}
\label{Seq:DeltaT}
\frac{ \Delta T(T_1, E_2, E_1) }{T_1} =  \frac{ 3 \zeta a^2 }{8 \pi }  \left[ \xi^{-2}(T_2, E_2) - \xi^{-2}(T_1, E_1) \right].
\end{align}
where $\Delta T =  T_2 - T_1$ 
 and we have made
 the assumption that the changes in temperature are small ($  \frac{\Delta T}{T_1}  \ll 1 $).  
This corresponds to  Eq. (\ref{eq:DeltaT}) in the main paper. 

To derive our figure of merit, we evaluate Eqs.~(\ref{Seq:DeltaS}) and~(\ref{Seq:DeltaT}) at the transition temperature for
 $E_2=E_0>0$ and $E_1=0$. 
From Eqs. ~(\ref{eq:mft_eq_RFE_1})-(\ref{eq:corrlength})
one can show that at $T_c^0$, 
\begin{align}
 \frac{\xi^{-2} (T_c^0, E_0 ) }{a^{-2}}  &= \frac{4 \pi v_0}{\zeta k_B C_{CW}  } p^2(T_c^0,E_0),
\end{align}
where $ p(T_c^0, E_0) = \left( k_B C_{CW} E_0 / v_0^2 \right)^{1/3} $. The saturated polarization $P_s =  n e^* p(T=0,E_0=0) $ 
is also derived from Eqs.~(\ref{eq:mft_eq_RFE_1})-(\ref{eq:mft_eq_RFE_4}) and the result is $P_s =  \sqrt{9 k_B n T_c^0 / (4 \pi) }$. By substituting all these results
into  Eq.~(\ref{Seq:DeltaS})  we obtain,
\begin{align*}
\frac{ \left| \Delta S ( T_c^0, E_0 ) \right| }{N k_B} 
= \frac{1}{2}\left( \frac{27}{4\pi} \right)^{2/3}  \left( \frac{ T_c^0 }{ C_{CW} } \right)^{1/3} \left( \frac{E_0}{P_s^0} \right)^{2/3},
\end{align*}
where we have done the  rescaling $ E_0 \to E_0 e^* $ so that $E_0$ in the above equation has units of electric field (see Hamiltonian).  The same result is obtained for $  \Delta T ( T_c^0, E_0 )/T_c^0$. This is our figure of merit given in Eq. (\ref{eq:fom}) of the main text.

\end{document}